\documentclass[conference]{IEEEtran}
\IEEEoverridecommandlockouts
\usepackage{cite}
\usepackage{algorithm}
\usepackage{algpseudocode}
\usepackage{diagbox}
\usepackage{amsmath,amssymb,amsfonts,amsthm}
\usepackage{breqn}
\usepackage{graphicx}
\usepackage{textcomp}
\usepackage{xcolor}
\usepackage{tabularx}
\usepackage{amsmath}
\usepackage[justification=raggedright,singlelinecheck=false]{caption}
\usepackage{booktabs} 

\def\BibTeX{{\rm B\kern-.05em{\sc i\kern-.025em b}\kern-.08em
    T\kern-.1667em\lower.7ex\hbox{E}\kern-.125emX}}
    
\begin{document}

\title{A Task Equalization Allocation Algorithm Incorporating Blocking Estimation and Resource Similarity Analysis for Vehicle Control Real-Time Systems
}
\author{\IEEEauthorblockN{Qianlong duan} 
\IEEEauthorblockA{\textit{School of Transportation Science and Engineering} \\ \textit{Beihang University}
\\ Beijing, China 
\\ duanqianlong@buaa.edu.cn} 
\and 
\IEEEauthorblockN{Bide Hao} 
\IEEEauthorblockA{\textit{School of Transportation Science and Engineering} \\ \textit{Beihang University}
\\ Beijing, China 
\\ haobide2000@buaa.edu.cn}
\and 
\IEEEauthorblockN{Shichun Yang}
\IEEEauthorblockA{\textit{School of Transportation Science and Engineering} \\ \textit{Beihang University}
\\ Beijing, China \\
yangshichun@buaa.edu.cn} 
\and 
\IEEEauthorblockN{Fei Chen} 
\IEEEauthorblockA{\textit{School of Transportation Science and Engineering} \\ \textit{Beihang University}
\\ Beijing, China
\\ johnfei1@126.com}
\and \IEEEauthorblockN{Fan Zhou*} 
\IEEEauthorblockA{\textit{School of Transportation Science and Engineering} \\ \textit{Beihang University}
\\ Beijing, China 
\\ fanzhou@buaa.edu.cn} }

\maketitle

\begin{abstract}
In multi-core real-time vehicle control systems, synchronization blocking and resource contention pose critical challenges due to increasing task parallelism and shared resource access. These issues significantly degrade system schedulability and real-time performance, as traditional task allocation algorithms often overlook blocking impacts, leading to high scheduling failure rates under heavy loads. To address this, we propose the BR-WFD algorithm, which integrates blocking time estimation and resource similarity analysis. The algorithm minimizes global blocking overhead by prioritizing tasks with high synchronization sensitivity and aggregating shared-resource-accessing tasks onto the same core. Extensive simulations show that BR-WFD reduces required processor cores by 11\% to 28\% and maintains a 15\% to 20\% higher schedulable ratio compared to traditional methods under high-load and resource-competitive scenarios. This demonstrates its effectiveness in enhancing real-time performance and resource efficiency for multi-core task scheduling in intelligent driving systems.
\end{abstract}


\section{Introduction}
With the rapid advancement of intelligent connected vehicle technology, the complexity of tasks in vehicle-mounted control systems is continuously escalating, involving numerous control and perception tasks that require concurrent execution. This high parallelism imposes higher requirements on computing resources, making multi-core processor platforms a requirement for meeting deadlines. Within this context, achieving efficient collaborative scheduling of tasks on multi-core platforms has become a core challenge in system design~\cite{sheikh2018energy,zhang2019energy,ansari2019peak,suyyagh2019energy,behera2020energy}.

In intelligent driving systems, vehicle control tasks have strict real-time constraints and exhibit complex resource access dependency relationships. As computing platforms evolve from single-core to multi-core architectures, the frequency of concurrent task scheduling and shared resource access has significantly increased, leading to increasingly prominent synchronization blocking issues caused by resource contention. Most traditional task allocation algorithms are designed primarily with computational load balancing as the core concept, failing to consider the impact of synchronization blocking on system performance entirely. In scenarios with high load or intense resource competition, this oversight can significantly increase system scheduling failure rates, severely affecting task response times and system stability. Therefore, there is an urgent need for a new strategy that can proactively consider resource conflicts during the task scheduling phase to enhance system scheduling performance further.

For the real-time task scheduling problem involving shared resource access, Nemati et al. proposed a clustering strategy based on the Best Fit Decreasing (BFD) heuristic to partition task sets accessing shared resources \cite{nemati2010partitioning}. Additionally, Tsai et al. proposed an algorithm based on the Multiprocessor Stack Resource Policy (MSRP) to enhance system energy efficiency by limiting task synchronization \cite{tsai2015triple}. This algorithm focuses on adjusting execution frequency and reclaiming dynamic idle time to reduce energy consumption, but does not deeply explore how to optimize real-time schedulability through task partitioning.
To address the above challenges, this paper focuses on synchronization blocking issues caused by shared resource access in multi-core systems. We construct a task worst-case response time (WCRT) analysis model to accurately evaluate the impact of different blocking types on task response times. Based on this, we propose a task-balanced allocation algorithm named BR-WFD (Blocking Time Estimation and Resource-awareness Worst-Fit Decreasing) that integrates three key technical advantages:

1) Blocking-Time-Driven Prioritization: By precomputing each task’s blocking time estimation utilization (PBU), the algorithm identifies tasks with high synchronization sensitivity and schedules them first, reducing their exposure to cross-core blocking risks.

2) Resource-Aware Aggregation: Leveraging resource similarity analysis, the algorithm co-locates tasks accessing the same resources onto the same core. This minimizes global resource contention, as local resource access under the Multiprocessor Priority Ceiling Protocol (MPCP) avoids cross-core priority inversion and remote blocking.

3) Adaptive Load Balancing: Using a hybrid worst-fit strategy, the algorithm dynamically selects cores with the smallest blocking load (BU) when initial resource-similar cores are overloaded. This prevents load imbalance-induced scheduling failures while maintaining real-time guarantees.   By incorporating these mechanisms, BR-WFD achieves a $30\%$ reduction in average blocking time compared to traditional approaches and a $25\%$ enhancement in core utilization efficiency. Simulation experiments confirm that the algorithm exhibits superior system schedulability and achieves $20\%~\sim25\%$ faster task response times across various critical section configurations. These results underscore its significance as a reliable solution for real-time task scheduling.

The remaining structure of this paper is as follows: Section~\ref{sect:ref} overviews related work on resource access protocols and task allocation algorithms; Section~\ref{sect:system_modeling} analyzes task models and task schedulability; Section~\ref{PropsoedAlgo} introduces the main components of the proposed BR-WFD algorithm; Section~\ref{sect:EVALUATION} presents the empirical performance of the proposed BR-WFD algorithm; and Section~\ref{sect:conclusion} summarizes the paper and our main findings.

\section{Related Work}
\label{sect:ref}

This work is most related to Resource Access Protocols and Task Scheduling Algorithms, which will be introduced in the following sub-sections.

\subsection{Resource Access Protocols}
Due to the limited nature of system resources in typical embedded systems, multiple tasks inevitably compete for shared resources during execution, often requiring management through exclusive access. In environments with concurrent multi-task execution, frequent resource access conflicts may cause task blocking, thereby increasing response times and degrading the overall real-time performance of the system. Therefore, designing efficient resource-sharing protocols is fundamental to achieving multi-task allocation and scheduling, which is crucial for improving system schedulability and operational stability.

In real-time operating systems, semaphores are commonly used by tasks to protect critical sections that access shared resources, while the actual management of the resources is handled either by the operating system or cooperatively by the tasks. Before entering a critical section, a task must acquire the semaphore associated with the corresponding resource. In many real-time application scenarios, the need for functions to exclusively access shared resources often leads to priority inversion \cite{sha1990priority}, particularly on multi-core computing platforms.

Researchers have proposed various resource access protocols to address resource access contention and the resulting priority inversion. In single-processor systems, for fixed-priority scheduling, Sha et al. proposed the Priority Ceiling Protocol (PCP) \cite{sha1990priority}. For dynamic priority scheduling (such as EDF), Baker proposed the Stack Resource Policy (SRP) \cite{baker1991stack}. Subsequent research extended these single-processor platform resource access protocols to multi-processor platforms, such as the Multiprocessor Priority Ceiling Protocol (MPCP) \cite{lakshmanan2009coordinated} and the Multiprocessor Stack Resource Policy (MSRP) \cite{gai2003comparison}. Additionally, studies have proposed more flexible multi-processor locking protocols, such as the Flexible Multiprocessor Locking Protocol (FMLP) \cite{block2007flexible} and the Suspension-based Optimal Locking Protocol (OMLP) \cite{brandenburg2010optimality}. Recently, for fixed-priority partitioning scheduling problems, scholars have proposed the Multiprocessor Resource Sharing Protocol (MrsP) and conducted feasibility analyses \cite{burns2013schedulability}. Furthermore, with the widespread deployment of mixed-criticality systems, research on resource access protocols for multi-criticality task synchronization has received increasing attention, with relevant literature deeply exploring protocol adaptability and real-time guarantees from perspectives such as response time analysis and security isolation mechanisms \cite{swiecicka2006multiprocessor, zhao2015resource}.

\subsection{Task Scheduling Algorithms}
Traditional task scheduling algorithms commonly used in embedded real-time systems can be broadly divided into two categories: static-priority scheduling and dynamic-priority scheduling. Static scheduling algorithms are further classified into hybrid-priority and fixed-priority approaches, depending on whether task priorities may change \cite{verucchi2023survey}. In real-time systems, Fixed-Priority (FP) scheduling is a widely used paradigm. Among FP algorithms, two classical examples are Rate-Monotonic (RM) and Deadline-Monotonic (DM), both of which assign task priorities according to timing parameters. Specifically, RM assigns higher priorities to tasks with shorter periods, while DM assigns higher priorities to tasks with shorter relative deadlines. Both are preemptive scheduling models, allowing higher-priority tasks to preempt lower-priority ones during execution.

Dynamic scheduling algorithms include the Least Laxity First (LLF) scheduling algorithm and the Earliest Deadline First (EDF) scheduling algorithm. In the EDF scheduling algorithm, task priorities are determined by their deadlines, with the system prioritizing tasks with the earliest deadlines to ensure the most urgent tasks are executed first. The algorithm dynamically adjusts task execution order to ensure the task with the earliest deadline is executed at each moment. While dynamic scheduling algorithms are particularly effective in improving schedulability and deadline adherence, especially in systems with diverse tasks and tight timing constraints, they may also lead to starvation of long-running tasks under high system loads.

With the development of multi-core processor technology, the research focus of task scheduling has gradually shifted to how to effectively map tasks to different processor cores and solve resource access contention and task synchronization issues in multi-core systems. In fact, real-time task scheduling in multi-core processor systems can be abstracted as combinatorial optimization problems similar to the Travelling Salesman Problem or the Knapsack Problem \cite{salman2022multi}. Researchers have introduced various constraints, such as reliability, energy consumption, and development costs, to optimize scheduling algorithms further \cite{hu2023online, deng2021reliability, huang2020dynamic}. Researchers have proposed various optimization methods to address different application scenarios, such as meta-heuristic algorithms and deep reinforcement learning algorithms, for task scheduling optimization in multi-core systems.

Building on this, Han et al. proposed a synchronization-aware Worst-Fit Decreasing (WFD) task partitioning algorithm, SA-WFD (Synchronization-Aware Worst-Fit Decreasing) \cite{han2012synchronization}. This algorithm aims to partition tasks requiring access to the same resources onto the same processor core to improve schedulable ratios. Saad et al. extended the SA-WFD concept to heterogeneous multi-core systems \cite{saad2013energy}. 

\section{System Model Building}
\label{sect:system_modeling}
\subsection{Task Model}
This paper focuses on hard real-time task scheduling in vehicle control system environments, where all tasks must strictly complete before their deadlines. Task allocation algorithms involve various task models, among which periodic tasks are prevalent in vehicle control applications, with typical period sets widely adopted in research and industrial practice, including $\{1, 2, 5, 10, 20, 50, 100, 200, 1000\}ms$\cite{hamann2017communication,kramer2015real,sailer2013optimizing,tobuschat2016system, von2017parametric}. Therefore, the task synchronization allocation algorithm proposed in this paper primarily targets periodic tasks, with the following detailed introduction to the basic definitions and modeling of periodic task models.

A periodic task $i$ is represented by a quadruple $\tau_i = (C_i, T_i, D_i, \Pi_i)$, where $C_i$ is the task's worst-case execution time, $T_i$ is the task's execution period, $D_i$ is the task's deadline, and $\Pi_i$ is the task's execution priority—higher $\Pi_i$ indicates higher execution priority. 
The periodic task allocation problem in multi-core processors can be formally described as follows: given a periodic task set $\Gamma={\tau_1,\tau_2,\dots,\tau_n}$, this set will be deployed on a processor architecture $P=\{p_1,p_2,\dots,p_m\}$ with $m$ computing cores. A bidirectional mapping relationship can be established between $\Gamma$ and $P$ as follows,
 \begin{itemize}
     \item[(1)]Core location function: $p_{\tau_i} \to p_j$ indicates the processor core where task $\tau_i$  resides.
     \item[(2)]Task allocation function: $\tau(p_k)\subseteq\Gamma$ indicates the subset of tasks hosted by the processor core $p_k$.
 \end{itemize}

It is assumed that the system contains $q$ shared resources $\varphi=\{r_1,r_2,\dots,r_q\}$, where $\Theta_i \subset \varphi$ denotes the set of shared resources accessed by $\tau_i$. 
Resource access follows the mutual exclusion principle: at any time, a resource $r_s \in \varphi$ can be held by at most one task. 
Furthermore, tasks are assumed to access resources in a non-nested manner, i.e., each task $\tau_i$ may hold at most one resource in $\Theta_i$ at a time. 
$t_{i,s}$ represents the time that $\tau_i$ occupies $r_s$, \emph{i.e.}, the critical section execution time.

Shared resources can be divided into local resources and global resources. If $r_s\in \varphi$ can only be accessed by tasks allocated to the same core in $\Gamma$, then $r_s$ is a local resource; otherwise, it is a global resource. If a task $\tau_i$ does not access any shared resources with other tasks, it is called an independent task. $lp(i)$ denotes the set of tasks on the processor core where $\tau_i$ resides, with priorities lower than $\tau_i$, and $hp(i)$ denotes the set of tasks on the same core with priorities higher than $\tau_i$.
Let $L$ be the least common multiple of all periodic tasks, \emph{i.e.,} the hyper-period of the task set. It is generally assumed that if all tasks are schedulable within one hyper-period, the task set is globally schedulable. Within one hyper-period $L$, a task with period $T_i$ will be scheduled $L/{T_i}$ times. The task model studied in this paper adopts the implicit deadline assumption, \emph{i.e.,} all tasks' deadlines are equivalent to their periods.

\subsection{Blocking Model}
In multi-core real-time systems, concurrent access to shared resources by tasks can trigger issues such as priority inversion and deadlock, severely affecting system predictability and real-time performance. To address the problem of mutual exclusion access to shared resources by tasks in multi-processor environments, this paper employs the widely used Multiprocessor Priority Ceiling Protocol (MPCP) as the resource access protocol.

The core idea of the MPCP is to raise the task priority to the resource's "priority ceiling" during resource access to prevent low-priority tasks from blocking high-priority tasks due to holding critical resources. In multi-core environments, MPCP not only manages local processor core resource mutual exclusion but also coordinates cross-core task access to global shared resources through global priority elevation and scheduling synchronization mechanisms, ensuring the system maintains good scheduling predictability and analyzability while sustaining high concurrency. Assume $\tau_h$ is the highest-priority task accessing shared resource $r_k$. 
Any task executing within the critical section of shared resource $r_k$ has its priority elevated to the ceiling priority $\Omega_k=\Pi_b+\Pi_h$, where $\Pi_b$ is a higher priority than any normal task, and $\Pi_h$ is the execution priority of $\tau_h$. MPCP includes the following properties:
\begin{itemize}
    \item A task executes at its base priority when it is not inside a critical section. Upon entering a critical section, its priority is elevated to the corresponding ceiling level: local resources are governed by PCP, while global resource requests follow the MPCP mechanism.
    \item If task $\tau_i$ acquires a global resource $r_k$, it temporarily elevates to the ceiling priority $\Omega_k$, corresponding to that resource. However, it may be preempted by another task $\tau_j$ accessing a resource $r_x$ with a higher ceiling priority $\Omega_x > \Omega_k$.
    \item If a global resource $r_k$ is not currently occupied by another task, a task's request is immediately granted. Otherwise, the task is added to a priority-ordered waiting queue, sorted by the task's original priority.
    \item When a task releases a global resource $r_k$, if there are tasks waiting for this resource, the highest-priority task in the queue is granted access. If no tasks are waiting, the resource is released directly.
\end{itemize}

Although MPCP provides good real-time guarantees and analyzability in multi-core systems, its synchronization mechanisms can still cause scheduling interference in high shared-resource contention scenarios. The following analyzes typical synchronization blocking types in MPCP.

\begin{itemize}
    \item \emph{Transitive Remote Preemption} is a preemptive phenomenon that indirectly prolongs task blocking. The transitive remote preemption blocking process is shown in Fig.~\ref{fig:frame1}. When task $\tau_x$ running on core $p_a$ is waiting for a global shared resource $r_k$, when $r_k$ is currently held by task $\tau_x$ on core $p_b$ within its critical section, if task $\tau_y$ on core $p_b$ accesses a global resource $r_l$ with a higher ceiling priority, it will preempt the execution of $\tau_x$ according to MPCP rules ($\Omega_l>\Omega_k$), thereby prolonging the waiting time of $\tau_i$.
    \begin{figure}
    \centering
    \includegraphics[width=0.9\linewidth]{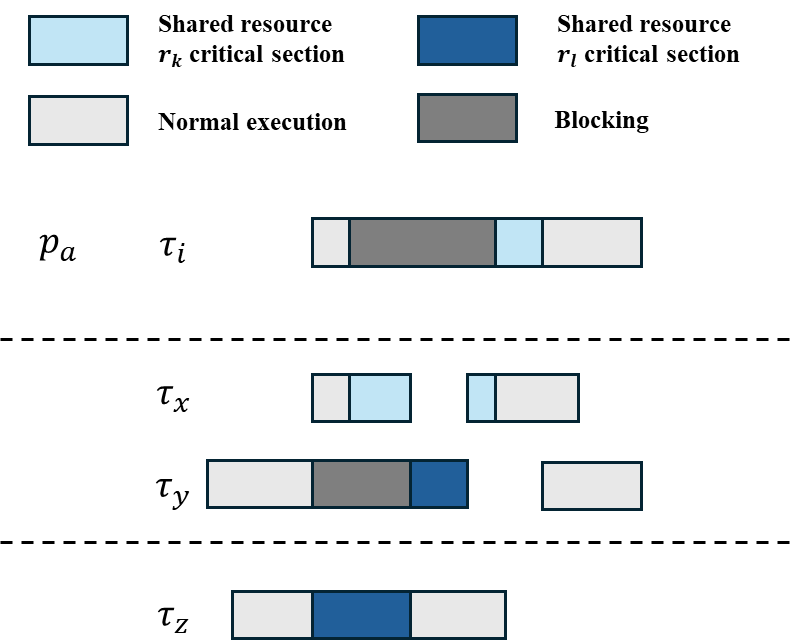}
    \caption{Transitive Remote Preemption}
    \label{fig:frame1}
\end{figure}

    \item \emph{Multiple Remote Blocking} refers to a phenomenon that, when a task requests a global shared resource, it may be blocked not only by the current resource holder but also by multiple higher-priority tasks on other cores. In MPCP, resources are controlled by a global priority queue, where higher-priority tasks acquire resources first. Thus, even if a task has waited for some time, it must continue queuing as long as higher-priority tasks request the resource. The multiple remote blocking process is shown in Fig.~\ref{fig:frame2}. Assuming task priorities $\Pi_z > \Pi_y > \Pi_x > \Pi_y$, task $\tau_i$ is first remotely blocked by task $\tau_z$. During the execution of $\tau_z$, tasks $\tau_x$ and $\tau_y$ on core $p_b$ are activated and access shared resource $r_k$. Due to their higher priorities, $\tau_i$ is further blocked by $\tau_x$ and $\tau_y$. In summary, task $\tau_i$ undergoes continuous remote blocking from multiple higher-priority tasks due to accessing the shared resource $r_k$.
    
    \begin{figure}
        \centering
        \includegraphics[width=0.9\linewidth]{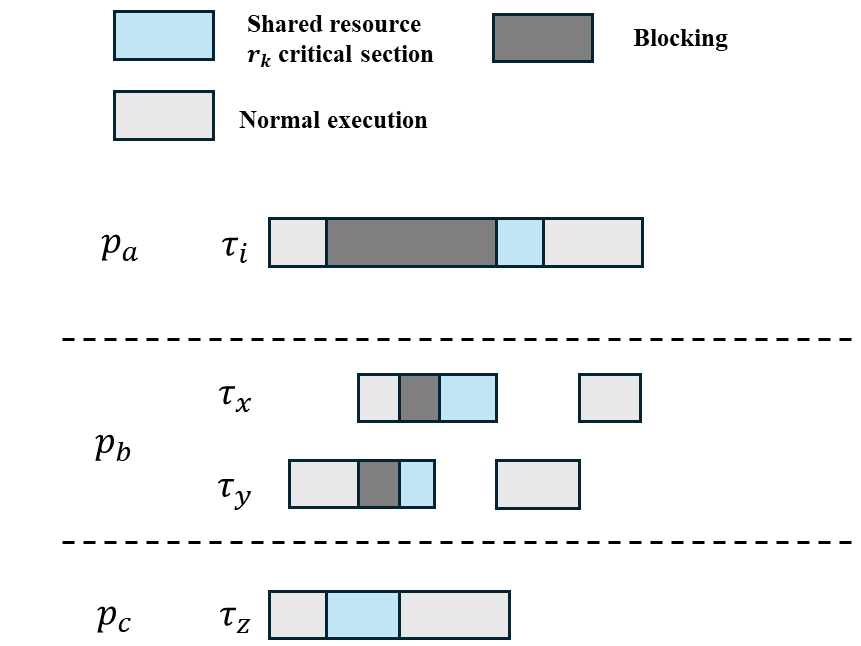}
        \caption{Multiple Remote Blocking }
        \label{fig:frame2}
    \end{figure}

    \item \emph{Multiple Priority Inversion}. This is an issue where low-priority tasks repeatedly affect high-priority task execution. When a high-priority task is suspended due to waiting for a resource, the system scheduler may allocate the processor to ready low-priority tasks, allowing them to execute before the high-priority task. According to the MPCP operation, the high-priority task can only resume normal scheduling after the shared resource is released. However, during this recovery interval, other low-priority tasks may re-enter the run queue and occupy the processor, causing the high-priority task to be delayed multiple times, forming a cascading priority inversion phenomenon. The multiple priority inversion process is shown in Fig.~\ref{fig:frame3}. 
    \begin{figure}
    \centering
    \includegraphics[width=0.9\linewidth]{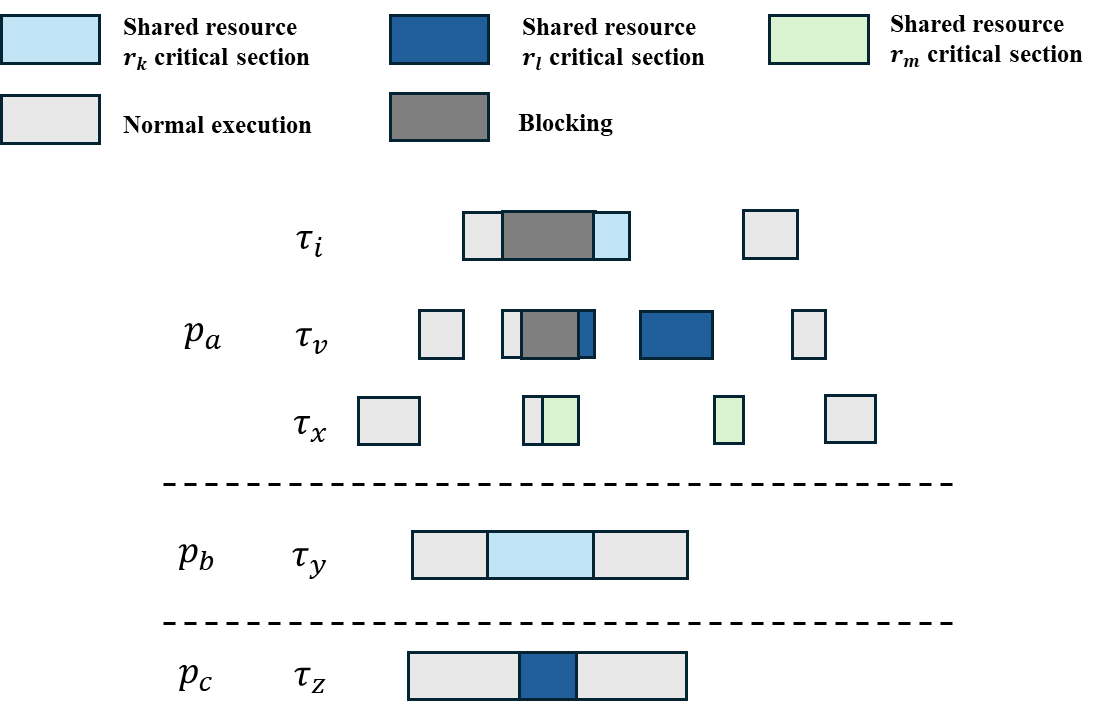}
    \caption{Multiple Priority Inversion}
    \label{fig:frame3}
\end{figure}
    Assuming task priorities $\Pi_z > \Pi_y > \Pi_i > \Pi_v > \Pi_x$, and ceiling priorities $\Omega_k > \Omega_l > \Omega_m$, task $\tau_i$ is blocked by task $\tau_y$ on core $p_b$ while accessing shared resource $r_k$. During the blocking of $\tau_i$, task $\tau_V$ on core $p_b$ is scheduled but is blocked by task $\tau_z$ when attempting to access shared resource $r_l$. Subsequently, low-priority task $\tau_x$ is scheduled on core $p_a$ and accesses shared resource $r_m$. Since $\tau_v$ and $\tau_x$ hold resources $r_l$ and $r_m$, respectively, after releasing the shared resource $r_k$ and resuming normal execution, task $\tau_i$ will still be preempted by $\tau_v$ and $\tau_x$ in sequence—despite their lower priorities—triggering a chain of multiple priority inversion phenomena that significantly impact system real-time guarantees. 
    
\end{itemize}

\section{Proposed Algorithm}
\label{PropsoedAlgo}

Synchronization blocking under the MPCP primarily originates from blocking mechanisms such as transitive remote preemption, multiple remote blocking, and multiple priority inversion. These blocking interferences can significantly prolong task blocking times during resource access, thereby affecting their response times and overall system schedulability. To quantitatively analyze their impact, the following models and upper-bound analyses of task synchronization blocking times are conducted from the perspective of worst-case response time, referencing the synchronization analysis method proposed by Yang et al. \cite{yang2013synchronization}.

First, consider task \(\tau_{i}\) running on processor core \(p_{a}\), blocked by a global shared resource \(r_{k}\) currently locked by another task \(\tau_{j}\) running on a different processor core \(p_{a}\). When tasks on \(p_{b}\) access global resources with higher ceiling priorities, they may preempt task \(\tau_{j}\)’s execution, indirectly causing \(\tau_{i}\) to be remotely and transitively preempted. The upper bound of this transitive remote preemption impact can be determined by the formula:

\begin{equation}
\label{equation1}
\alpha_{j, k}=\sum_{\forall \tau_{u} \in p\left(\tau_{j}\right)} \max _{\forall r_{x} \in \Theta_{j} \wedge \Omega_{k}<\Omega_{x}} \gamma_{u, x}^{\max }
\end{equation}

where \(\gamma_{u,x}^{max}\) is the maximum duration task \(\tau_{u}\) accesses shared resource \(r_{x}\) in a single instance.

Next, analyze different sources of blocking that task  \(\tau_{i}\) may encounter.

1) Local Resource Blocking: During task \(\tau_{i}\)’s suspension, low-priority tasks may request local resources, delaying \(\tau_{i}\)’s resumption. Since \(\tau_{i}\) can be suspended at most \(N_{i,G}\) times, the upper bound of local resource blocking time can be estimated by Equation:

\begin{equation}
\label{equation2}
DLB_{i}=
\begin{pmatrix}
1+N_{i,G}
\end{pmatrix}
\max_{\substack{
\forall\tau_{j}\in lp(i)\cap p(\tau_{i}) \\
\wedge r_{l}\in\Theta_{j}\cap\Theta_{L}\wedge\pi_{i}<\Omega_{l}
}}\gamma_{j,l}^{\max}
\end{equation}

where \(N_{i,G}\) is the number of critical sections of global shared resources accessed by task \(\tau_{i}\),  and \(\Theta_{L}\) is the set of local shared resources.

2) Global Resource Low-Priority Blocking: When task \(\tau_{i}\) requests a global resource, it may be blocked by tasks running on other processor cores, potentially causing transitive remote preemption. For blocking by low-priority tasks, each request for a global resource \(r_k\) by  \(\tau_{i}\) results in at most one block. Therefore, the upper bound of direct global resource blocking time from low-priority tasks can be estimated by Equation (3):

\begin{equation}
\label{equation3}
DGB_{i}^{L}=\sum_{r_{k}\in\Theta_{i}\cap\Theta_{G}}N_{i,k}\cdot\max_{\forall\tau_{j}\in lp(i)\cap\tau\left(\tilde{p}(\tau_{i})\right)}\left(\gamma_{j,k}^{\max}+\alpha_{j,k}\right)
\end{equation}

where \(\Theta_{G}\) is the set of global shared resources, and \(N_{i,k}\) is the number of critical sections where task \(\tau_{i}\) accesses shared resource \(r_k\).

3) Global Resource High-Priority Blocking: Considering the impact of multiple remote blocking on task \(\tau_{i}\), whenever  \(\tau_{i}\) queues for a global resource \(r_k\) higher-priority tasks running on other processor cores requesting this resource will block \(\tau_{i}\). Moreover, these higher-priority tasks may execute multiple times during \(\tau_{i}\)’s waiting period.  Therefore, the direct global resource blocking time caused by higher-priority tasks can be upper-bounded by Equation (4):

\begin{equation}
\label{equation4}
\begin{split}
DGB_{i}^{H} = \sum_{r_{k}\in\Theta_{l}\cap\Theta_{G}} &\sum_{\forall\tau_{j}\in hp(i)\cap\tau\left(\widetilde{p}(\tau_{l})\right)} \\
&\left[\frac{T_{i}}{T_{j}}\right] \cdot \left(\gamma_{j,k}^{\mathrm{total}} + N_{j,k} \cdot \alpha_{j,k}\right)
\end{split}
\end{equation}

where \(\gamma_{j,k}^{\mathrm{total}}\) is the total duration task \(\tau_{j}\)spends accessing shared resource \(r_k\) during execution.

4) Local Priority Inversion Blocking: This blocking term arises from multiple priority inversion interferences. When task \(\tau_{i}\) is suspended, low-priority tasks on the same processor 
\(p(\tau_{i})\) may start running and queue for global resource access, potentially causing priority inversion and interrupting \(\tau_{i}\)’s normal execution. During \(\tau_{i}\) execution, each low-priority task \(\tau_{j}\) can initiate at most \(2N_{j,G}\) requests for global resources, with blocking time upper-bounded by Equation (5):

{\small
\begin{equation}
\label{equation5}
MLI_{i}=\sum_{\forall\tau_{j}\in lp(i)\cap p(\tau_{l})}\min\left(1+N_{i,G},2N_{j,G}\right)\mathrm{max}_{r_{k}\in\Theta_{j}\cap\Theta_{G}}\gamma_{j,k}^{\mathrm{max}}
\end{equation}
}

Finally, under the MPCP, the worst-case blocking time (WCBT) for task \(\tau_{i}\)  is the sum of the above blocking times:

\begin{equation}
\label{equation6}
B_{i}=DLB_{i}+DGB_{i}^{L}+DGB_{i}^{H}+MLI_{i}
\end{equation}

Blocking caused by synchronization interference delays task execution. This additional interference in response time can be upper-bounded by the maximum remote blocking time. Therefore, the worst-case response time for task \(\tau_{i}\) can be calculated using the following convergence formula:

{\small
\begin{equation}
\label{equation7}
W_{i}^{n+1}=C_{i}+B_{i}+\sum_{\forall\tau_{j}\in hp(i)\cap p(\tau_{i})}\left\lceil\frac{W_{i}^{n}+DGB_{i}^{H}+DGB_{i}^{L}}{T_{j}}\right\rceil\cdot C_{j}
\end{equation}
}

The initial value for this iteration is:

\begin{equation}
\label{equation8}
W_{i}^{0}=C_{i}+D G B_{i}^{H}+D G B_{i}^{L}
\end{equation}

The iteration continues until convergence:

\begin{equation}
\label{equation9}
W_i^{n+1}=W_i^n
\end{equation}

If the result of an iteration exceeds the task's relative deadline, it indicates the task cannot complete on time, and the iteration can be terminated early to determine scheduling failure.

Based on the analysis of blocking time sources in Equations (2) to (5), global blocking terms induced by task shared resource access dominate the total blocking time. These global blockages are not only closely related to task priorities and resource access order but also trigger synchronization blocking interference on remote processor cores in multi-core environments, significantly increasing task worst-case response times and directly affecting system real-time performance and schedulability.

To reduce synchronization blocking and improve task scheduling success rates, this paper proposes a task balanced allocation algorithm based on blocking time estimation and resource awareness (BR-WFD), which builds on the following observations: In multi-core systems, blocking time caused by global resources often has a more significant impact on overall task response times and is a primary factor leading to task allocation failures. Therefore, BR-WFD estimates global blocking times for tasks based on their priorities and shared resource allocation status before allocation, prioritizing tasks with higher global blocking impacts to reduce overall system scheduling pressure. Meanwhile, to further minimize inter-task resource contention, the algorithm fully considers task-to-task resource correlation during task mapping, striving to allocate tasks accessing the same shared resources to the same processor core, thereby reducing global blocking caused by global resource access conflicts. This strategy aims to enhance task set schedulability at the system level by reducing global blocking time. The BR-WFD algorithm primarily includes three core features: 

1) Blocking Time Estimation-Based Sorting Mechanism: Before task allocation, calculate each task's blocking time estimation utilization and sort tasks in descending order, prioritizing tasks with higher blocking time estimation utilization.

2) Resource-Aware Task Aggregation Strategy: During task allocation, preferentially map tasks accessing the same shared resources to the same processor core to reduce global blocking time.

3) Load Balancing Strategy: When a target processor core is overloaded, use the WFD strategy to allocate tasks to the processor core with the smallest current blocking load, ensuring overall system load balance and reducing task allocation failures.

The overall execution flow of the BR-WFD algorithm (Algorithm 1) is as follows: Taking the task set to be allocated and the processor set as input, the algorithm outputs the task allocation results \(H^j\) for each processor core after execution. The allocation algorithm is broadly divided into the following phases:

Phase 1: Input the unallocated task set \(\Gamma\) and the processor set P,where each task \(\tau_{i}\) in \(\Gamma\) should include at least task priority \(\tau_{i}\), execution time \(C_i\), execution period \(T_i\), accessed shared resource set \(\Theta_i\) and critical section execution times and counts.

Phase 2: In lines 1-3 of the algorithm, calculate the utilization of blocking time estimation of each task \(\tau_{i}\) using Equation (12) and sort tasks in descending order of utilization of blocking time estimation.

Phase 3: Tasks are allocated sequentially according to the precomputed order. For each task, the algorithm calculates its resource similarity with each processor core, defined as the sum of overlaps in accessed shared resources (Equation 15). The task is initially assigned to the core with the highest similarity. However, if this allocation results in a blocking load (BU) exceeding the current maximum BU across all cores, the algorithm switches to the core with the lowest BU to ensure load balancing. After allocation, parameters such as core utilization and blocking load are updated. A Response Time Analysis (RTA) is then performed using Equations (7)-(9) to verify schedulability. If the task set fails the RTA, the algorithm terminates immediately, indicating an unschedulable configuration.

Phase 4: Output the task set  \(H_j\) and utilization  \(U_j\) for each processor core.

\begin{algorithm}
\caption{Task Allocation Algorithm Based on Blocking Time Estimation and Resource Awareness (BR-WFD)}\label{alg:cap}
\begin{algorithmic}[1]
\Require  \(\Gamma\): Unallocated task set,  P: Processor core set \({p_1, p_2, ..., p_m}\)
\Ensure \(H^j\): task set allocated to each processor core
\State 	Initialize task set \(H^j \leftarrow \emptyset\), utilization \(U^j \leftarrow 0\), blocking load \({BU}^j \leftarrow 0\) for each core;
\State 	For each task \(\tau_{i}\) , calculate its blocking time estimation utilization \({PBU}_{i}\) using Equation (12);
\State Sort all tasks in descending order of \({PBU}_{i}\);
\For{each sorted task \(\tau_{i}\)} 	
    \State Calculate the resource similarity   between \(\tau_{i}\)  and each processor core; 
    \State Select the processor core with the highest resource similarity as the initial candidate core;
    \State If allocating \(\tau_{i}\)  to this core would cause \({BU}^j\) to exceed the current maximum blocking load, select the core with the smallest current \({BU}^j\) using the worst-fit strategy;
    \State Allocate \(\tau_{i}\)  to the selected processor core and update \(H^j\), \({BU}^j\), and \(U^j\) accordingly; 
    \State If \({BU}^j\) exceeds the current maximum blocking load, update the maximum blocking load;
    \State Perform schedulability analysis for the current allocation; if the task set is unschedulable, terminate the algorithm and return an empty allocation set;
\EndFor
\State Repeat until all tasks are allocated;
\State Return the final task allocation \(H^j\) and utilization  \(U^j\) for each core.
\end{algorithmic}
\end{algorithm}

To quantitatively evaluate blocking impacts, the BR-WFD algorithm introduces a task blocking time estimation mechanism. This mechanism models and estimates the blocking impact on task \(\tau_{i}\) from both low-priority and high-priority perspectives, incorporating blocking factors from global resource access:

First, based on Equation (3), assuming all shared resources accessed by \(\tau_{i}\) are global and all low-priority tasks may cause global blocking, the estimated blocking time for \(\tau_{i}\) from low-priority tasks is:

\begin{equation}
\label{equation10}
PGB_i^L=\sum_{r_k\in\Theta_i}\max_{\forall\tau_j\in l(i)\wedge r_k\in\Theta_j}\gamma_{j,k}^{\max}
\end{equation}

Where \(l(i)\) denotes the set of all tasks in the task set with priorities lower than \(\tau_{i}\).

Further, based on Equation (4), the estimated blocking time for task \(\tau_{i}\) from high-priority tasks is:

\begin{equation}
\label{equation11}
\begin{split}
PGB_i^H = \sum_{r_k\in\Theta_i} &\sum_{\forall\tau_j\in h(i) \wedge r_k\in\Theta_j} \left[\frac{T_i}{T_j}\right] \gamma_{j,k}^{\mathrm{total}}
\end{split}
\end{equation}

Where h(i) denotes the set of all tasks in the task set with priorities higher than \(\tau_{i}\).

Based on these two blocking estimates and the task's own execution time \(C_i\), the blocking time estimation utilization (PBU) for task \(\tau_{i}\) can be further calculated:

\begin{equation}
\label{equation12}
PBU_i=\frac{C_i+\beta(PGB_i^L+PGB_i^H)}{T_i}
\end{equation}

The PBU metric evaluates task load and blocking intensity during the task sorting phase, where \(\beta\) is a proportionality coefficient used to adjust the blocking time estimation weight, adjustable according to the order-of-magnitude ratio between critical section execution time and task execution time.

On this basis, the blocking load on the current processor core \(P_j\) is defined as the sum of the blocking time estimation utilizations of all allocated tasks:

\begin{equation}
\label{equation13}
BU^j=\sum_{\forall\tau_i\in p_j}PBU_i
\end{equation}

This value serves as an important reference for measuring processor core scheduling pressure and is used in the BR-WFD algorithm's load balancing strategy to select target processor cores with the lightest blocking load, achieving reasonable task mapping and scheduling optimization in multi-core systems.

During task allocation, preferentially mapping tasks to processor cores with similar resource access types not only enhances resource access locality but also significantly reduces global blocking time generated when tasks request shared resources, thereby improving overall system schedulability and operational efficiency. To quantify inter-task resource access similarity, this paper uses a resource correlation coefficient \(\omega_{i,j}\) to measure the overlap in shared resource types between tasks \(\tau_{i}\) and \(\tau_{i}\), defined as:

\begin{equation}
\label{equation14}
\omega_{i,j}=|\Theta_i\cap\Theta_j|
\end{equation}

Based on inter-task resource correlation, the resource similarity \(\phi_{i}^{x}\) between task \(\tau_{i}\) and processor core \(p_x\) is further defined, representing the sum of resource correlation coefficients between \(\tau_{i}\) and all allocated tasks on processor core \(p_x\) when task \(\tau_{i}\) is allocated to \(p_x\), with the specific calculation formula:

\begin{equation}
\label{equation15}
\phi_i^x=\sum_{\forall\tau_j\in\tau(p_\chi)}\omega_{i,j}
\end{equation}

\section{PERFORMANCE EVALUATION}
\label{sect:EVALUATION}

\subsection{Experimental Methods and Parameter Configuration}

To verify the performance of the BR-WFD allocation algorithm, this paper conducted extensive simulation experiments with default parameter settings as shown in Table I. Task execution times \(C_i\) were randomly generated using a uniform distribution within the interval [20,100] ms, referencing the execution duration distribution of periodic tasks in typical vehicle control systems. Task utilization \(u_i\) was generated using the UUnifast algorithm within the interval [0.1,0.15]. According to the definition of task periods in real-time systems, period \(T_i\) was derived from the formula \(T_i=C_i/u_i\). Each task included 2 to 3 critical sections, with no correlation between shared resources accessed in different critical sections. The RM scheduling algorithm was used for task scheduling, so task priorities were inversely proportional to their execution periods.

To simulate task access patterns to shared resources in multi-task systems, shared resources were divided into several resource groups, with each group containing 5 shared resources by default. The system allocated tasks to access one group of shared resources per 15 tasks. In each round of experiments, 1,000 task sets were randomly generated, and the task allocation performance of scheduling algorithms was evaluated under different settings. Due to the small order-of-magnitude difference between task execution times and critical section execution times, the proportionality coefficient \(\beta\) in Equation (12) was set to 0.1. Unless otherwise specified, simulation experiment parameters used default values. System total load was defined as the sum of task utilizations for all tasks in a single task set, and the critical section execution time for shared resources accessed by tasks was the product of task execution time and the critical section ratio.

\begin{table}[htbp]\large
\caption{Simulation Experiment Parameter Configuration}
\begin{center}
\begin{tabular}{c c}
    \hline
    Parameter & Default Value/Range \\ \hline
    Task execution time (ms) & [20,100]\\ 
    Task utilization & [0.1,0.15]\\ 
    Number of shared resources & [2,3] \\
    Critical section ratio & [0.12] \\
    System total load & [8] \\
    \hline
\end{tabular}    
\end{center}
\end{table}
\subsection{Performance Metrics}

The following performance metrics were used in simulation experiments to evaluate algorithm:

(1) Number of Processor Cores Required: Defined as the minimum number of processor cores needed to complete task set allocation and scheduling. The task allocation process first attempts to map the task set to a number of processor cores equal to the system total load. If the number of processor cores is insufficient for allocation, the number of cores is gradually increased, and allocation is re-executed until task allocation is successfully completed or deemed failed. A smaller number of required processor cores indicates lower system resource requirements for the algorithm, demonstrating better adaptability and resource utilization efficiency.

(2) Schedulable Ratio: This metric represents the percentage of task sets meeting feasibility test conditions relative to the total number of tested task sets. A higher schedulable ratio indicates stronger scheduling capability of the partitioning algorithm, enabling effective operation in a broader range of application scenarios. Therefore, as a core performance metric for real-time scheduling algorithms, the schedulable ratio can effectively evaluate the Pros and Cons of partitioning algorithms in real-time schedulability, serving as a key consideration in real-time scheduling research.

\subsection{Experimental Result Analysis}

\begin{figure}
    \centering
    \includegraphics[width=1.0\linewidth]{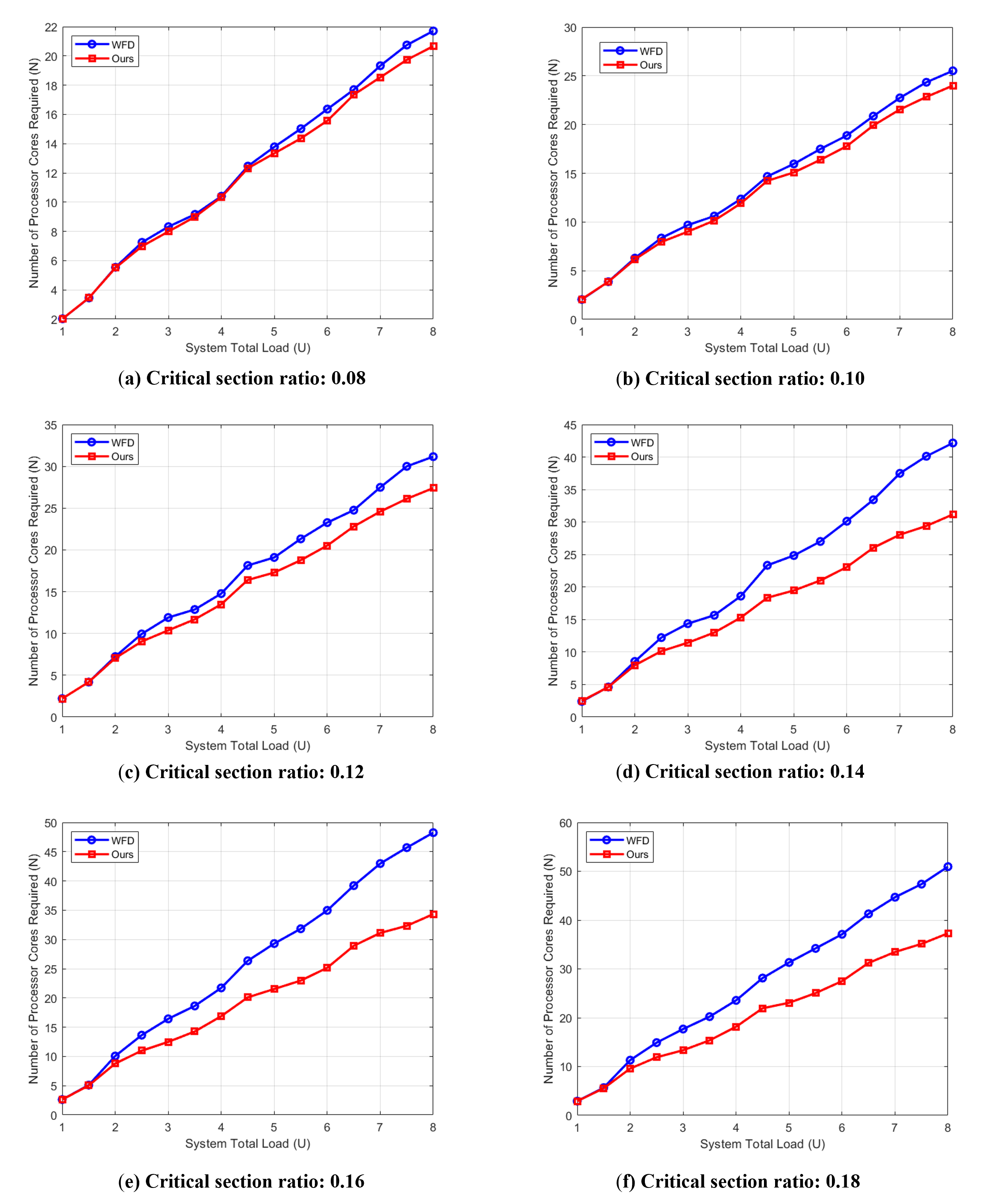}
    \caption{Effect of different critical section ratio on cores number}
    \label{fig:frame4}
\end{figure}

Fig.~\ref{fig:frame4} illustrates the combined influence of system total load and critical section ratio on the number of required processor cores across varying critical section ratios, while Table 2 further quantifies the reduction amplitude of processor cores achieved by the proposed BR-WFD algorithm compared to the baseline WFD algorithm. Experimental results reveal that under high-load conditions (S, system total load $>$ 6) combined with intense resource competition (C, critical section ratio $>$ 0.12), the BR-WFD algorithm significantly reduces the number of required processor cores by 11\% to 28\% compared to the WFD algorithm, demonstrating its superior efficiency in resource-constrained scenarios. This result indicates that in complex scenarios with increased scheduling pressure and frequent synchronization conflicts, the BR-WFD algorithm can effectively mitigate task synchronization blocking times, improve task response speeds, and ensure tasks complete before their deadlines. Compared to the WFD algorithm, BR-WFD demonstrates superior schedulability with the same number of processor cores, thereby enhancing processor resource utilization and overall system efficiency, further validating its advantages and practical application value in multi-core task scheduling. From the perspective of overall system performance, the BR-WFD algorithm not only ensures timely task completion but also improves the overall throughput and energy efficiency of multi-core systems. It demonstrates excellent scalability and practical applicability in complex real-time scheduling scenarios.

In summary, the experimental results validate the significant performance gains of the BR-WFD algorithm under extreme scheduling conditions and highlight its potential for task scheduling in multicore processors. The algorithm exhibits clear advantages in reducing resource waste, enhancing system efficiency, and improving scheduling robustness, thereby offering strong support for the design and optimization of complex real-time systems.

\begin{figure}
    \centering
    \includegraphics[width=1.0\linewidth]{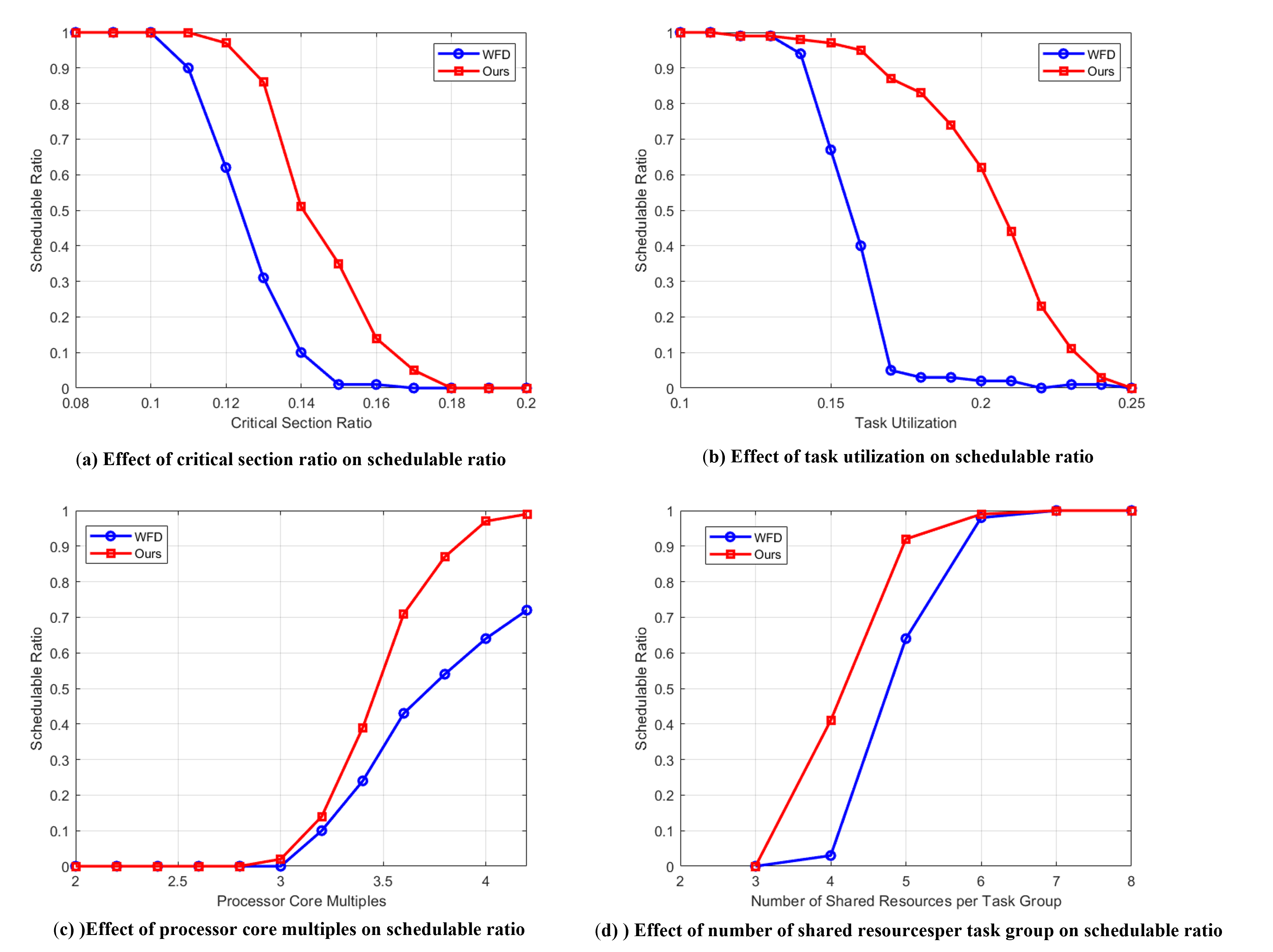}
    \caption{Effect of different parameters on the schedulable ratio}
    \label{fig:frame5}
\end{figure}

Fig.~\ref{fig:frame5} systematically demonstrates the impact of multiple key parameters on task set schedulable ratios:

Fig.~\ref{fig:frame5}(a) shows the impact of critical section ratio on schedulable ratios. With other parameters held constant, a larger critical section ratio means tasks occupy shared resources for longer durations during execution, intensifying competition and conflict over shared resources in the system and reducing the likelihood of tasks completing on time. Experimental results show that when the critical section ratio exceeds 0.1, the schedulable ratio of the WFD algorithm significantly declines; in contrast, the BR-WFD algorithm exhibits stronger robustness under the same conditions, with a more gradual decline in schedulable ratio. Especially when the critical section ratio is in the 0.1–0.15 range, BR-WFD consistently maintains significantly better scheduling performance than WFD, demonstrating its more stable real-time scheduling capability in high-resource-competition environments.


\begin{table}[]
\caption{The optimization percentage of the required number of processor cores}
\renewcommand{\arraystretch}{1.5}
\centering
\begin{tabular}{@{}c|cccccc@{}}

\hline
  \diagbox{C}{S}& 0.08   & 0.1    & 0.12    & 0.14    & 0.16   & 0.18  \\ \hline
1 & 0.20\% & 2.05\% & 0.09\%  & 2.91\%  & 0.53\%  & 0.27\%\\
2 & 0.61\% & 2.19\% & 2.57\%  & 6.89\%  & 12.77\% & 15.24\%\\
3 & 3.90\% & 7.04\% & 12.88\% & 20.44\% & 24.01\% & 24.55\%\\
4 & 0.54\% & 3.66\% & 8.69\%  & 17.74\% & 22.25\% & 22.93\%\\
5 & 3.21\% & 5.51\% & 9.47\%  & 21.65\% & 26.49\% & 26.37\%\\
6 & 4.85\% & 5.77\% & 11.81\% & 23.39\% & 28.05\% & 25.86\%\\
7 & 4.10\% & 5.26\% & 10.54\% & 25.20\% & 27.59\% & 25.09\%\\ 
8 & 4.79\% & 6.01\% & 12.08\% & 26.08\% & 28.87\% & 26.80\%\\ \hline
\end{tabular}
\end{table}

Fig.~\ref{fig:frame5}(b) shows the impact of task utilization on schedulable ratios. With other parameters fixed, increased task utilization implies less idle time per task, increasing the risk that blocking causes tasks to exceed their deadlines. Experimental results indicate that when task utilization exceeds 0.14, the schedulable ratio of the WFD algorithm declines rapidly, while the BR-WFD algorithm exhibits a more gradual decline, demonstrating better scheduling elasticity. Notably, when task utilization reaches 0.17, the WFD algorithm is almost completely unschedulable, whereas the BR-WFD algorithm still maintains a schedulable ratio of approximately 0.95, showcasing significant scheduling performance advantages.

Fig.~\ref{fig:frame5}(c) shows the impact of processor core multiples on schedulable ratios. Processor core multiples are defined as the ratio of the system total load to available processor cores, with larger values indicating more abundant system resources. Significant scheduling performance improvements are only observed in high-load and intense-resource-competition environments when the number of processor cores exceeds three. In contrast, the BR-WFD algorithm achieves faster schedulable ratio improvements as core multiples increase, approaching a schedulable ratio of 1 when multiples exceed 4; the WFD algorithm, however, only maintains schedulable ratios between 0.6 and 0.7 under the same conditions. This indicates that BR-WFD can more efficiently utilize system resources to achieve superior scheduling performance when resource allocation is relatively lenient.

Fig.~\ref{fig:frame5}(d) shows the impact of the number of shared resources per task group on schedulable ratios. With other conditions unchanged, reducing the number of shared resources intensifies inter-task resource access conflicts and increases synchronization blocking risks in the system. Experimental results show that when the number of shared resources is less than 6, the schedulable ratio of the WFD algorithm significantly declines; the BR-WFD algorithm, however, only exhibits similar performance degradation when the number of shared resources is less than 5, further verifying its scheduling robustness and resource adaptability in high-resource-competition scenarios.

\section{Discussion and Conclusion}
\label{sect:conclusion}

Aiming at the synchronization blocking problem in multi-core task flow collaborative scheduling for vehicle control systems, we propose a task allocation algorithm named BR-WFD that integrates blocking time estimation and resource awareness, based on a systematic analysis of shared resource competition mechanisms and their impact on task response times. 
The proposed algorithm introduces a synchronization blocking estimation mechanism during the task allocation phase and combines inter-task resource access similarity with load balancing strategies to effectively mitigate the negative impact of resource conflicts on scheduling feasibility at the allocation source.
To verify the algorithm's effectiveness, we conduct multiple groups of simulation experiments to systematically evaluate the scheduling performance of the BR-WFD algorithm under typical parameters such as different critical section ratios, task utilizations, processor core multiples, and shared resource configurations. 
Experimental results show that the BR-WFD algorithm significantly outperforms the traditional WFD algorithm in high-load and intense-competition environments. It effectively reduces the number of processor cores required for scheduling and maintains higher schedulable ratios under various adverse conditions, demonstrating good scalability and resource adaptability. The BR-WFD algorithm provides a practical technical path for synchronization-aware and efficient resource collaborative scheduling of real-time tasks in multi-core vehicle control systems, holding engineering application value for enhancing the real-time guarantee capability of complex vehicle control systems in harsh environments.
Furthermore, as the complexity of automotive control systems continues to increase, future multi-core automotive control systems will involve more types of shared resources, such as memory, bus, I/O devices, and others. The coordination and management of these resources will become more intricate. Achieving efficient scheduling and proper allocation across multiple resource types remains an unresolved challenge. Therefore, future work could consider integrating scheduling mechanisms for various resource types with the BR-WFD algorithm, developing a multi-resource collaborative scheduling framework to further enhance the overall scheduling performance and robustness of the system.

In conclusion, the BR-WFD algorithm provides an innovative solution for real-time task scheduling in automotive control systems, laying a solid foundation for the realization of efficient, low-latency, and stable multi-core automotive control systems. While the current work has made significant progress, further optimization and extension are necessary to address the increasingly complex and dynamic environments of future automotive control systems and ensure they meet the more stringent real-time requirements.

\bibliographystyle{IEEEtran} 
\bibliography{IEEEabrv,ref}


\end{document}